# Hypergamigication Through Integrating Game Engines and Learning Management Systems: Ender's Game

Araz Yusubov, Tangiz Alizada
School of IT and Engineering, ADA University
Baku, Azerbaijan
ayusubov@ada.edu.az

Michael Bechtel
K16 Solutions
Scottsdale, Arizona, USA
mike.bechtel@k16solutions.com

***Abstract* —This paper discusses games, their use in education, and previous work on integrating game engines and learning management systems (LMS). It proposes a bidirectional integration where game environments are generated using LMS content, introducing the concept of *hypergamification* as the use of a comprehensive game environment rather than isolated game design elements. A working pilot implementation of an importable *Unity* package for *Blackboard* integration is demonstrated, along with a demo game that uses the developed package. The paper also discusses the limitations of the proposed approach and outlines avenues for future work.**

## I. Introduction

Games represent a multifaceted cultural phenomenon, the social, evolutionary, developmental, art and educational dimensions of which have been researched extensively (Burghardt, 2012; Frenzel, 1977). They often appear connecting geographically and historically distant cultures. A curious example is the game known as *besh dash* among Turkic-speaking nations that can be found depicted in a 4th century BC ancient Greek terracotta statue, in a 16th century Dutch renaissance painting, in a 20th century Argentinean poem, and in a modern day Korea (Sajó et al., 2011). There is a whole interdisciplinary field called ludology that studies the games (Csikszentmihalyi, 1982; Frasca, 1999). The digital video games as a relatively novel phenomena represent an interesting topic as the video game industry is projected to reach $206.5 billion by 2028 (Global Games Market Report, 2025). A dedicated IEEE and ACM journal (IEEE Transactions on Games, ACM Games: Research and Practice) serves as a reputable source for the related academic research and industry practice.

There are a number of established formal frameworks for understanding games that include the MDA model (Hunicke et al., 2004), which introduces three main components of a game as mechanics, dynamics, and aesthetics, hence is the acronym. Mechanics represents primary elements from the designer's viewpoint, such as what players can or cannot do while trying to achieve the game goal. Aesthetics is about how the game looks, sounds, and feels, which represents the primary elements from the player's viewpoint. Finally, dynamics represents the "run-time behavior" as interplay of the player input and mechanics.

### A. *Games in Education*

Use of games in education, as a pedagogical method, a learning activity, etc., has been documented since the ancient times, and the relevant research, which is evident in the indexed databases since late 1960s, has boomed since 2010s (Shelton et al., 2011; Ekin et al., 2023). Several distinct concepts that emerged in the literature include game-based learning, serious games and gamification:

- Game-based learning is a pedagogical approach where to achieve the defined learning outcomes a specific game is used as a tool within an educational curriculum, and the learning happens as a direct consequence of playing the game. An example is using *Monopoly* to teach accounting.
- A serious game is a standalone game with educational content embedded into its design, so it is developed for a primary purpose, which is not pure entertainment. Examples are modern tactical war games or professional flight simulators.
- Gamification, in contrast, does not represent a game. It is defined (Deterding, 2012; Deterding et al., 2011) as the "use of game design elements in non-game contexts… to motivate desired behaviors." Getting points and badges for completing language learning tasks and progressing on the leaderboard in *Duolingo* is an example.

Video games have been used in education for many decades. Early serious games entered classrooms along with personal computers in 1980s' California (Juba, 2017), from *Oregon Trail* to the *Carmen Sandiego* series (Lin, 2024), which was rebooted in 2025. A commercial video game, such as *Spore*, is played as part of the class activities to provide context for reflecting on a curricular topic in biology (Rüth & Kaspar, 2021). There are even curious examples when during the global COVID-19 pandemic lockdown a popular video game, such as *Half-Life* (Bretan, 2020) or *Minecraft* (Nigmatulin, 2020), was effectively used as a virtual digital environment for learning.

### B. *Game Engines and Learning Management Systems*

Game engines emerged in the late 1980s as a component of the modern video game architecture responsible for the core mechanics such as underlying graphical systems and physics simulators. The most prominent of these proprietary software frameworks were developed in-house and reused to create successful game titles. In 2009, free versions of two major game engines, *Unity* and *Unreal Engine*, were released, boosting the production of high-quality games by independent developers.

Online Learning Management Systems (LMS) emerged in the late 1990s as centralized, web-based platforms that combine tools for learning content delivery, student progress tracking, communication, and administration (Coates et al., 2005; Pittinsky, 2025). The establishment of universal interoperability standards since the early 2000s allowed widespread institutional adoption. Over the years the modern LMS market matured into two distinct categories: Higher Education/Academia and Corporate Learning/Development. While transforming localized, computer-based training into interconnected digital environments, the modern LMS has evolved from a basic repository for course documents into an integrated cloud-hosted ecosystem.

Developing an integration module for these two systems would provide game designers with tools to create games that enrich the learners' experiences within their traditional LMS courses.

## II. Integrating Two Systems

To explore integration of an LMS with a game engine, we chose *Blackboard* and *Unity* as two popular products to experiment with. The choice of LMS was also informed by relevance to our host organizations at the time.

*Unity* remains a dominant game engine on the market as the leader by number of game titles published on the *Steam* platform (Video Game Insights, 2025). Capable of deploying content across 20+ platforms, including PC, mobile, consoles, and virtual reality devices, it uses C# as its primary scripting language and features robust graphics, physics, multiplayer support, a large user community and an extensive assets library.

*Blackboard*, one of the first LMSs released in 1997 and the most popular one on the North American higher education market back in 2009, still is among top four players along with *D2L Brightspace*, *Moodle*, and *Canvas* (Hill, 2025). In 2016, the cloud-based *Blackboard Learn Ultra* framework was launched to improve the user experience (versus the legacy *Original* course view) and to modernize its interface architecture. In 2021, it was acquired by *Anthology*, which aimed to integrate the LMS into an all-in-one software ecosystem, but filed for bankruptcy in late 2025. As of March 2026, the company successfully emerged from restructuring and was rebranded back to *Blackboard*. The current focus is the fully modernized *Ultra* deployment with features such as native AI course generation tools and enhanced analytics, while preparing for the total retirement of its *Original* courses.

### A. *Previous Work*

One of the prominent early works towards integrating an LMS and a video game was *SLOODLE* (Livingstone & Kemp, 2008; Livingstone et al., 2008), short for Simulation Linked Object-Oriented Dynamic Learning, which was implemented as a combination of scripted objects in the *Second Life* massive multiplayer online social platform and optional modules running on the *Moodle* LMS server. Initially it focused on providing a 3D virtual classroom that would represent the elements of a given course. Developed over the years, it lets students participate in mirrored course chats, update their course blogs, take multiple-choice quizzes and submit their assignments in the virtual world.

As *Second Life*'s hosting costs escalated, the project pivoted toward supporting *OpenSim*, its open-source self-hosting alternative. Ultimately, *SLOODLE* was abandoned, as its public code repositories have not been updated since 2014 (Edgar & Preibisch, 2018), due to the deprecation of its underlying *Adobe Flash* tools, its incompatibility with new versions of *Moodle* and the *PHP* core.

SCORM (Sharable Content Object Reference Model) is one of the interoperability standards that help to integrate with LMSs. The SCORM Integration Kit for *Unity* (Stals, 2021), which comes with a simple demo game project, provides essential functionality, such as basic runtime communication between the systems, progress and score tracking, student progress bookmarking, and a custom template to export the game project into a standard SCORM .zip package to be used within any supporting LMS.

This specific integration kit supports SCORM 1.2, while the framework progressed to newer versions, and is ultimately intended to be succeeded by the Experience API (application programming interface) standard (xAPI). Furthermore, the architectural philosophy of SCORM relies on self-contained course packages, so these modules mainly pass data out to LMS while reading back only minimal user data.

LTI (Learning Tools Interoperability) emerged as a later interoperability standard more suitable for the age of cloud services. It allows an LMS to establish a seamless, secure connection with an external web-based tool using token-based authentication protocols.

Several implementations of LTI-based *Unity* integration with an LMS have been documented. For example, (Winterhagen et al., 2020) presents a prototype of a serious game, which passes student player performance data to the *Moodle* system using the now deprecated LTI 2.0 version.

The latest LTI 1.3 version comes with the LTI Advantage package of services built on it to add new features that facilitate deeper integration via REST architecture.

REST (Representational State Transfer) architecture uses standard web protocols to enable communication with external systems. Every data object or functional component is considered an identifiable resource accessible via a unique

URL, called an endpoint, collectively forming an API. A RESTful system uses native HTTP methods (GET, POST, PUT, and DELETE) to perform operations, transferring small data payloads in JSON (JavaScript Object Notation) or XML (Extensible Markup Language) formats.

### B. *Integration Models*

Gamification and serious games, by definition, provide more avenues for technical integration. Most of the mentioned implementation examples can be considered a serious game, as the learning materials are embedded into the game mechanics, while the integration is limited to passing interpretation of player's in-game performance as the learner's level of achievement of the learning outcomes.

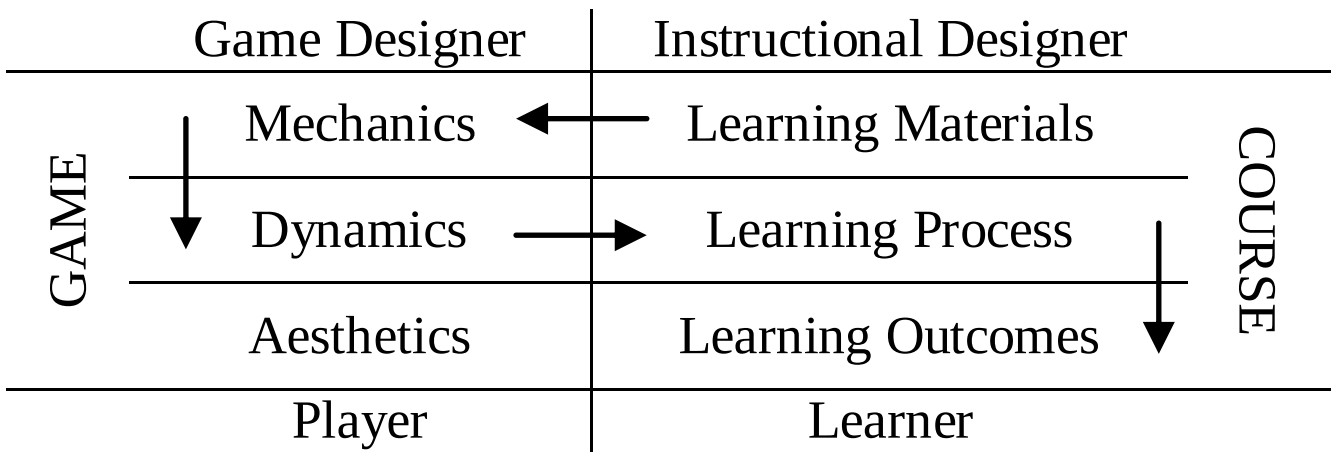

Figure 1. Parallels between game-related and educational components.

There is a curious similarity between game systems and learning systems. We can note that similar to how the MDA components are "causally linked," the learning materials, process, and outcomes are mutually interlinked (Fig. 1). These three educational components are in line with established learning frameworks, such as 3P (presage, process, product) (Biggs, 1987) and constructive alignment model (Biggs, 1996).

In a bidirectional integration model, learners will experience a virtual in-game environment that is generated based on the learning materials and course structure provided by the LMS. At the same time, interacting with this environment, playing the game will translate into learner activities in the course. This can be seen as *hypergamification* that will utilize not a few game design elements, but rather the entire game.

### C. *Gamification in Blackboard*

There is a considerable research in gamification of higher education (Khaldi et al., 2023; Lampropoulos et al., 2025), however LMSs in the academia category appear to lag behind the ones in the corporate learning. For example, *Blackboard* traditionally had a very few built-in features directly supporting gamification (Machajewski, 2017). For example (Yusubov, 2023), content modules can be used to group the curriculum materials under weekly learning outcomes like game levels, adaptive release can be used to release the next week's module only when minimal success is achieved in weekly self-check tests similar to unlocking game levels, achievement badges will be awarded for the test results, discussions post can be used as a leaderboard to manually record the weekly champions and announce the absolute champion at the end of the course.

It can be noted that these examples, as most of the early gamification implementations, are mainly centered around the PBL Triad of points, badges, and leaderboard. This design relies heavily on extrinsic motivation, where learners perform tasks primarily to earn external rewards.

The envisaged bidirectional integration promises to facilitate the implementation of alternative approaches, such as Octalysis Framework (Chou, 2012, 2019), which engage also more sustainable intrinsic motivation, where learners perform tasks because they appreciate the process and fulfill their personal desires.

### D. *Using Proprietary REST Protocols*

*Blackboard* natively supports both SCORM (versions 1.2 and 2004) and LTI (version 1.3 / Advantage). The legacy server-level Java framework, known as the proprietary Building Blocks (B2) standard, was completely retired as of June 2024. Currently, third-party integrations rely on LTI web services and *Blackboard*'s proprietary REST APIs, called Learn APIs. We used the latter for prototyping an integration module for *Unity*, which provides connectivity with *Blackboard*.

## III. Pilot Implementation

A proof of concept for the integration (Bechtel & Hart, 2022) successfully demonstrated connecting to *Blackboard*'s Learn API endpoints using the native `UnityWebRequest` class to obtain authorization tokens and access a course content item from within a *Unity* game. It highlighted the need for keeping the tokens refreshed once they expire, as well as for linking to the standard authentication workflows of the LMS. The demonstration also initiated discussions about the automated mapping of course objects into matching game objects.

The pilot implementation team was formed at ADA University's School of IT and Engineering, consisting of four senior undergraduate Information Technology students who continued the work as part of their capstone senior design project in 2023.

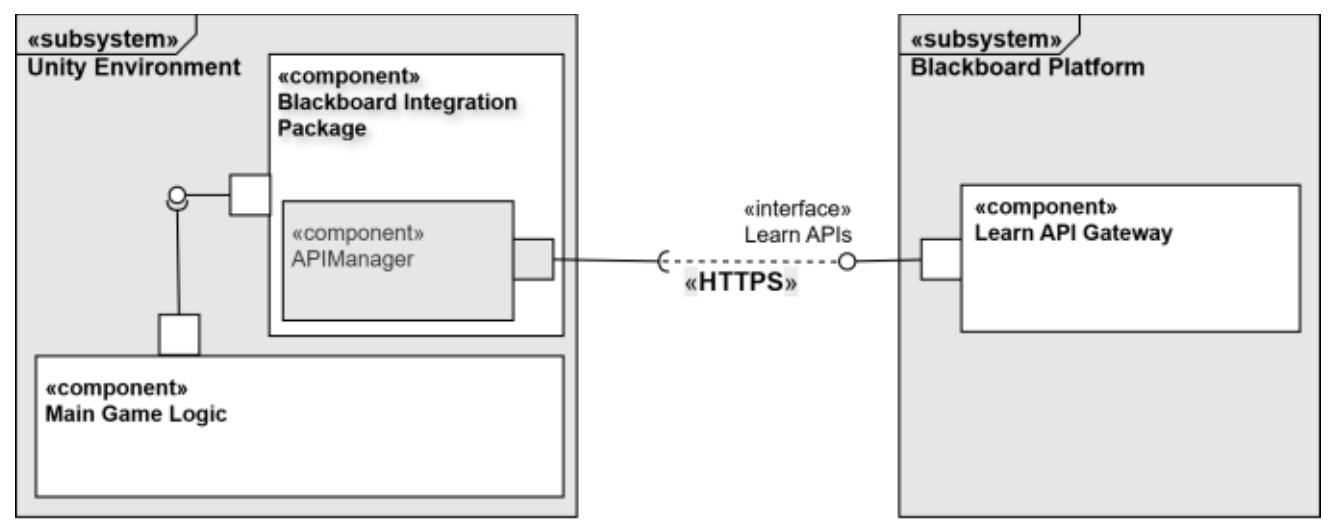

Figure 2. Components of the *Unity* to *Blackboard* integration.

To facilitate the experiments, the university provisioned a sandbox *Blackboard* environment, and provided an application key and a secret to acquire authorization tokens required for accessing the Learn API endpoints. The team started with exploring the APIs' capabilities by using *Postman* software environment, testing the retrieval of course data and making changes back in the course.

## A. Software Design

The two primary objectives of the project were to develop 1) a reusable *Unity* package (Fig. 2) that can be imported into game source code to conveniently use accessible *Blackboard* data and functionality, and 2) a demo video game to serve as a practical use case for this integration.

The software code developed for the pilot implementation is available in the public project repository (Alizada et al., 2023). The implementation was done using *Visual Studio 2022* and *Unity* version 2021.3.15f1. The version of *Blackboard*'s Learn APIs at the time was 3900.65.0.

The reusable *Unity* package contains two main script files: `APIManager.cs` and `SerializableItems.cs`. The former implements a singleton `APIManager` class that encapsulates all functionality required for LMS interoperability. It also contains critical properties, such as `CLIENT_ID` and `CLIENT_SECRET`, used to store application key and secret correspondingly.

```
public class APIManager : MonoBehaviour {
    public static APIManager Instance {
        get { return instance; }
    }
    private static APIManager instance;
    private const string CLIENT_ID = "xxx ";
    private const string CLIENT_SECRET = "xxx";
    private const string REDIRECT_URI = "unitydl://";
    private const string BASE_URL =
        "https://xxx.blackboard.com";

    private void Awake() {
        if (instance != null && instance != this) {
            Destroy(this.gameObject);
        } else {
            instance = this;
        }
    }
...
```

Figure 3. Part of the `APIManager` class implementation code.

The popular `Newtonsoft.Json` framework is used to handle serialization, i.e. converting local object states into web-friendly JSON payloads, as well as the reverse process of deserialization for incoming server responses.

Three methods manage the LMS authentication cycle. `GetAuthorizationCode()` invokes the native LMS login interface and, through *Unity* deep linking, redirects the user back to the game. Subsequently, `GetToken()` exchanges the obtained authorization code for an access token, while `GetNewToken()` refreshes the token to sustain secure communication.

There are twenty methods that connect to different Learn API endpoints to retrieve specific LMS object payloads and return them as deserialized class instances (Fig. 4). The required classes are declared in the `SerializableItems.cs` file. The access token obtained earlier during the authentication phase is passed to these asynchronous methods that make network requests without blocking the main *Unity* thread.

The demo video game was developed as a standalone application for the Universal Windows Platform (UWP). However, building it for the WebGL platform would be most suitable for achieving the sought bidirectional integration for hypergamification. These design choices, along with other limitations and challenges are discussed further in the text.

## B. In-Game User Experience

Once starting the demo game, users are presented with the school entrance scene (Fig. 5). There is a sphere in the center of the school yard that once clicked redirects the user to the *Blackboard* login page where they can enter their student login credentials.

Once logged in, the user returns to the game as an authorized student. This time, there are portals in the school yard for each available course. These portals, once approached, transport the student to the classroom scene.

```
public async Task<List<string>>
GetCourseIdsAsync(string user_uuid,
string access_token) {
    List<string> courseIds = new List<string>();
    using var www = UnityWebRequest.Get
    ($"{BASE_URL}/learn/api/public/v1/users/uuid:" +
    user_uuid + "/courses/");
    www.SetRequestHeader
    ("Authorization", $"Bearer {access_token}");
    www.SetRequestHeader
    ("Content-Type", "application/json");

    var operation = www.SendWebRequest();
    while (!operation.isDone) { await Task.Yield(); }
    if (www.result != UnityWebRequest.Result.Success) {
          Debug.Log($"Failed: {www.error}");
    }

    var jsonResponse = www.downloadHandler.text;
    try {
        MyCourses myCourses =
        JsonConvert.DeserializeObject<MyCourses>
        (jsonResponse);
        foreach (Course course in myCourses.results) {
            courseIds.Add(course.courseId);
        }
    } catch (Exception ex) { Debug.Log(ex.Message); }

    return courseIds;
}
```

Figure 4. An `APIManager` class method implementation.

On one of the walls in the classroom there is a big board with three main sections. The left section displays the student's announcements that are posted by instructors. The middle section displays the student's assignments, which may include instructions or file names. Clicking on an assignment automatically downloads it for the student. The right section displays the student's grades, including descriptions of the assignments, the grades earned, and the weighted total.

There is a row of tall cabinets under the big board, each representing a weekly materials folder. If the corresponding folder is not empty, clicking on a cabinet opens its door and displays the materials for that week. Clicking on a specific material automatically downloads it.

It can be noted that the choice of in-game 3D object representations of matching LMS objects for the demo game were informed by easily recognizable metaphors. However, we recognize that more creative matching would potentially produce a much more interesting experience.

### C. *Matching Objects*

It was observed that representation of different LMS elements as matching in-game objects is a separate topic to research. This mapping of objects done differently would result in different aesthetics, different user experiences. Perhaps the matching process can be standardized and automated based on, for example, a desired game genre (see Table 1).

TABLE I. SAMPLE MATCHING MAPS FOR DIFFERENT SCENARIOS

| LMS \ Game | Demo Game | A Role-Playing Game | A First-Person Shooter |
|---|---|---|---|
| **Student** | Player Student | Player Apprentice | Player Driver |
| **Institution** | School yard | Quest map | Headquarters |
| **Course** | Classroom | Quest | Operation |
| **Content Module** | Locker cabinet | Castle town | Deployment area |
| **Content Item** | Document | Commoner | Supply drops |
| **Assessment** | N/A | Guard | Field trials |
| **Announcements** | Board section | Market square | Radio traffic |
| **Announcement** | Board text | Town crier | Flash order |
| **Discussions** | N/A | Training yards | Garage floor |
| **Discussions Folder** | N/A | Training quarter | Vehicle station |
| **Discussion** | N/A | Training match | Maintenance log |
| **Gradebook** | Board section | Inventory | Service record |

To use elements of a role-playing game set in medieval Europe, the course could be represented as a quest world, where each content module is a castle town. The dwellers of these towns will correspond to content items of different types. The player can learn by ‘talking’ to them. Assessments would be guards chasing the player more actively as the deadline closes on. If it is a test assessment, the guard would start a dialogue and the player will choose or enter the answers.

### D. *Limitation and Challenges*

This pilot implementation project focused primarily on rapid prototype validation. Therefore, software optimizations and architectural refinements were deferred to future work.

Out of the 43 categories of endpoints testing was limited to OAuth, course announcements, attendance, content, content file attachments, content group assignments, course grades, course assessments, course membership, and users, mainly to GET method endpoints to retrieve data. The only `APIManager` method that sends data back to the LMS is `CreateMessage()`. It performs POST requests to send a message to the specified list of users in the given course.

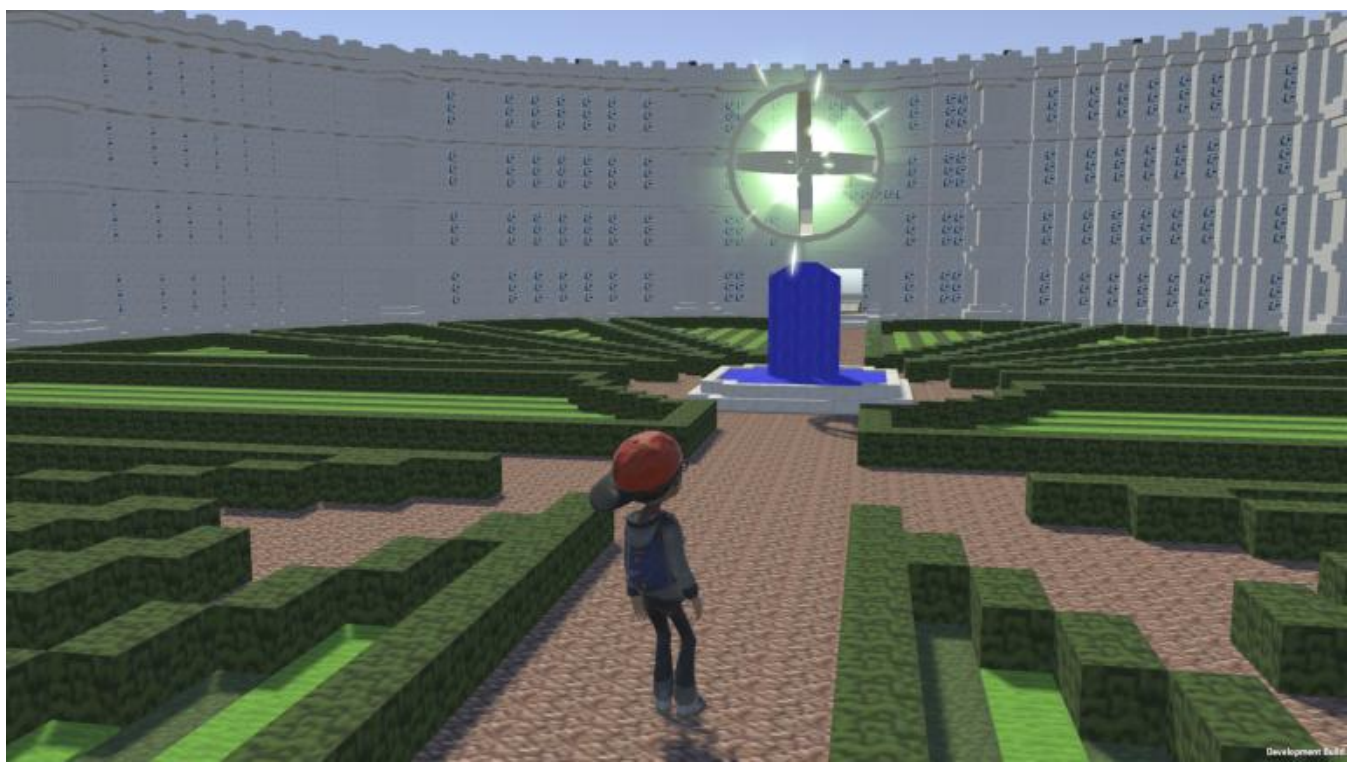

a

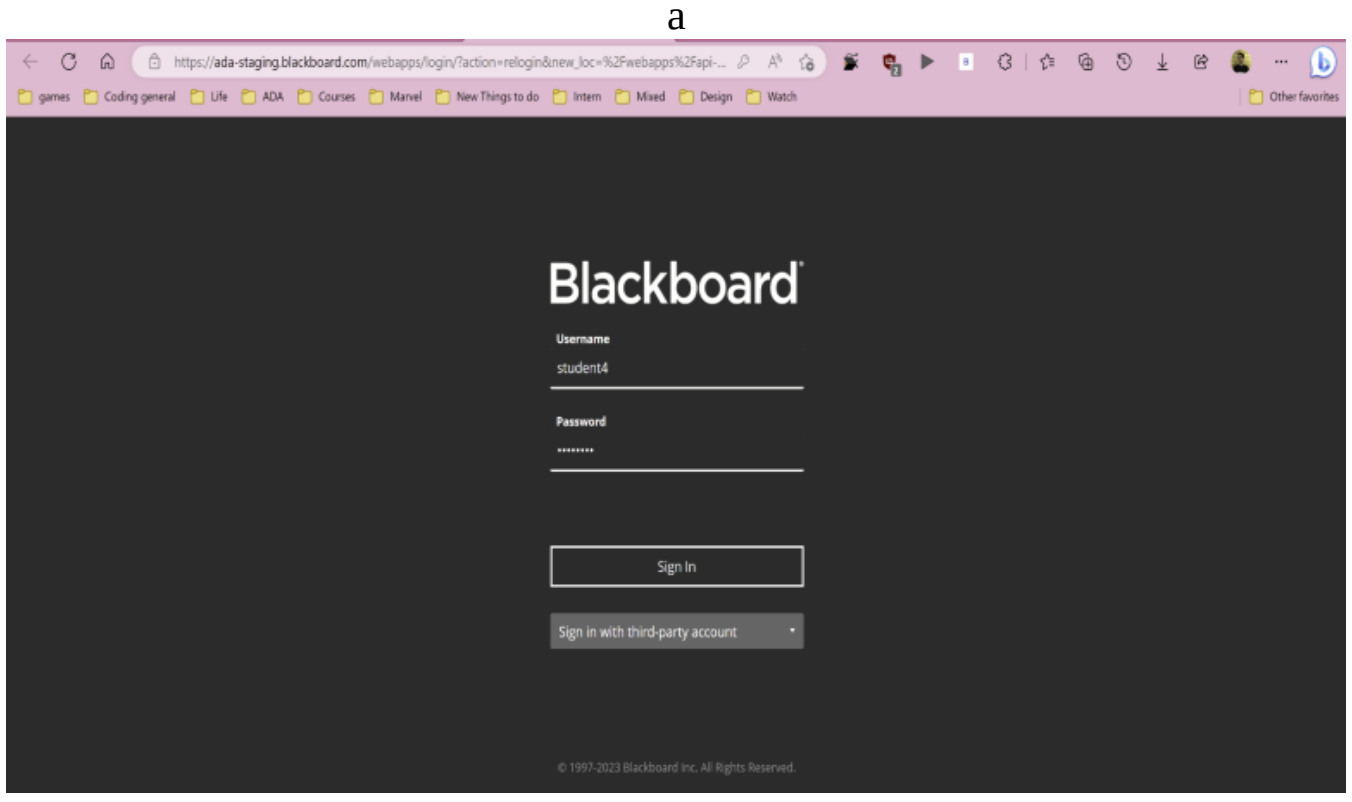

b

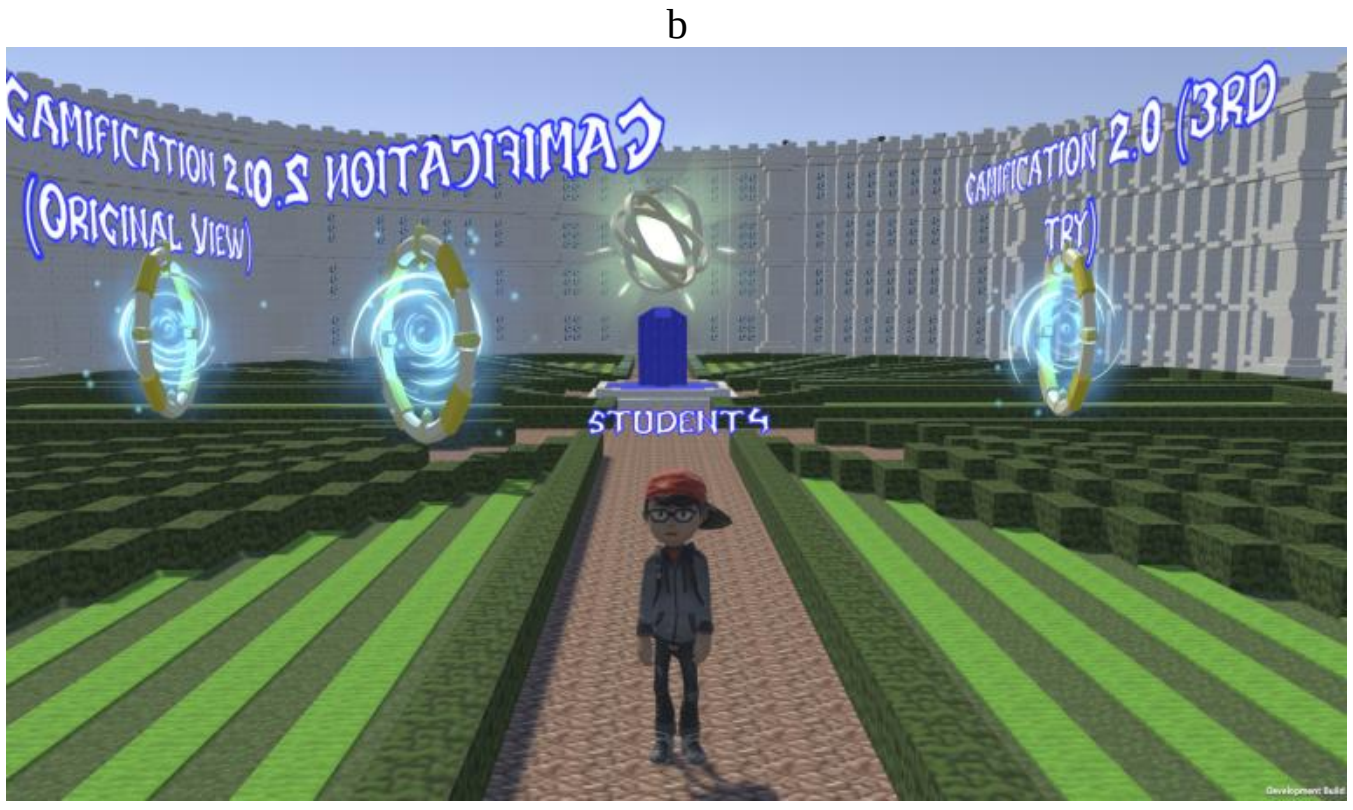

c

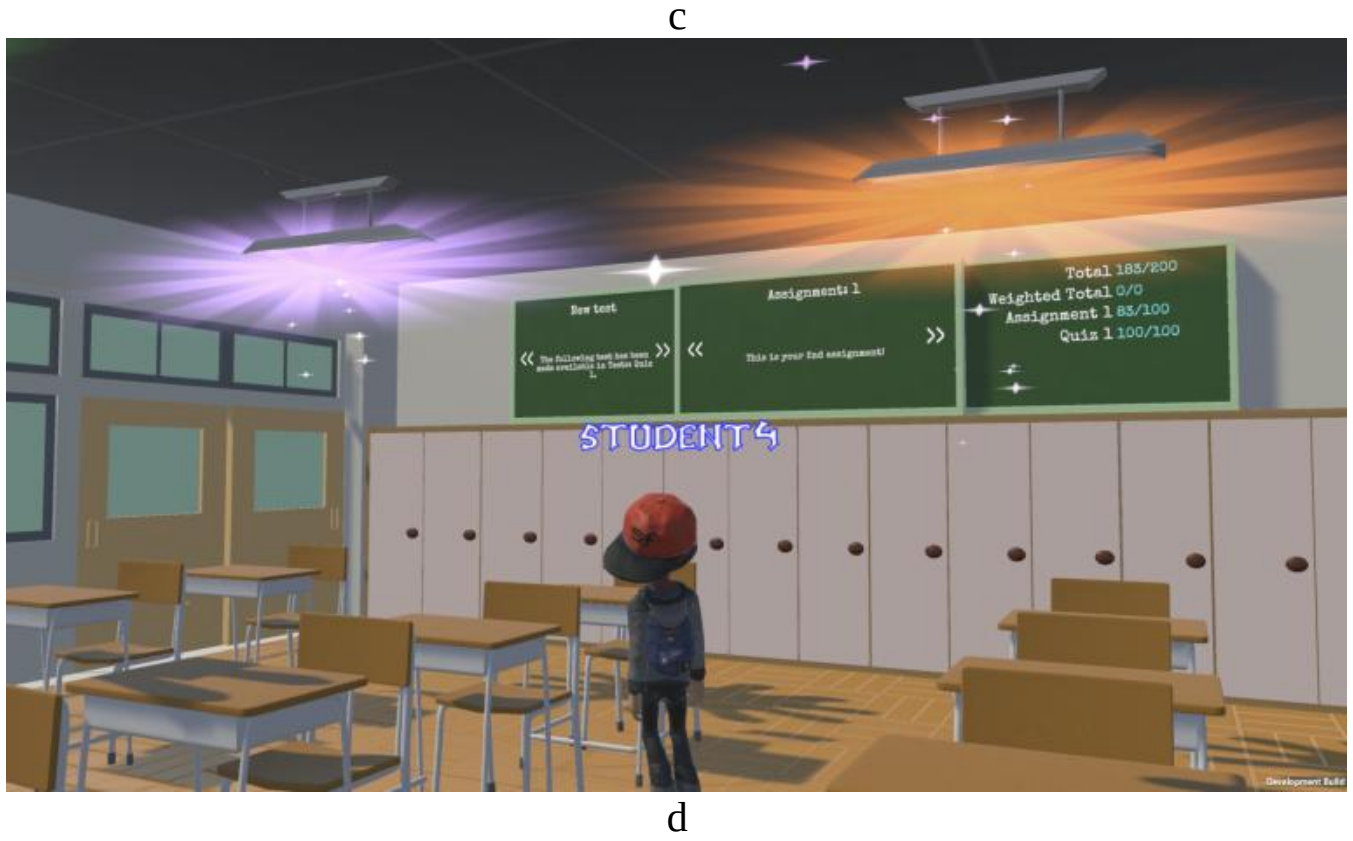

d

Figure 5. User experience in the demo game: (a) unauthorized student at the school yard (b) authorization through the LMS login (c) authorized student at the school yard (d) authorized student in the classroom.

The team tested retrieving the assessment details, including the questions and answer options, for implementing the use case mentioned earlier. The LMS response varied based on the authenticated user's role. For multiple-choice and open-ended questions the API gateway returns populated question schemas when an administrator or instructor-scoped token is used. Requests initiated with a student token returned no question content. Furthermore, there were no endpoints for posting runtime student answers. This limits the possibility of rendering assessments questions in a game environment.

In the pilot implementation the authorization uses *Unity* deep linking to redirect the user back to the game. For this prototype, the feature worked only if the game was built for the UWP platform. While UWP remains supported as a build target, it is considered a legacy framework and is not actively developed anymore.

## IV. Conclusion and Future Work

The paper explores interoperability of game engines and LMS systems. It presents integration of *Unity* and *Blackboard* through a pilot implementation project that produced a working prototype of an importable *Unity* package and a functional demo game. The project also demonstrated the limitations of the evaluated approaches and software frameworks.

With the working prototype, several future pathways emerge. This includes convening focus groups of educators and game developers to review the presented solution and brainstorm diverse use cases for more informed subsequent developments. For example, while the demo game does not utilize the multiplayer capabilities, exploring technical details of and models for using cooperative play would be interesting.

The existing codebase can be further optimized. For example, instead of passing the current access token to the `APIManager` class methods, it could be encapsulated as a private class property. Alongside individual methods retrieving the LMS data, a utility method to automatically traverse the hierarchy of *Blackboard* course objects and read relevant data would be useful. This method could be part of a separate class designed to encapsulate the student navigation within the course.

In scenarios where the Learn API cannot support the in-game rendering of specific LMS elements, third-party *Unity* plug-ins can be used to display and interact with the required LMS web pages inside the game environment.

Furthermore, *Blackboard* has introduced Ultra Extension Framework as a set of Premium APIs that use LTI 1.3 and REST to provide an authentication mechanism, allow the third-party applications to capture detailed user navigation and behavior data (e.g. click, route, and hover events), as well as to modify user interface elements and content. Leveraging this framework could help a more seamless integration of *Unity* games built for the WebGL target platform with *Blackboard*.

Finally, there is an emerging interest in using generative artificial intelligence for game development (Sharma et al., 2026; Humble, 2024). It would be interesting to explore methods to automate the generation of LMS course hypergamifications, personalized for diverse player preferences of the learners.

## Acknowledgments

The authors would like to thank Mr. Jonathan Hart, former Solutions Engineer at *Blackboard*, for useful comments and insights about the proof of concept, and the participants of the initial session at *Anthology Together 2022* conference for their lively interest. Special thanks go to Mr. Tural Hamzayev, Core Applications and Integrations Manager at ADA University, for arranging the sandbox *Blackboard* environment, as well as for unwavering support and steadfast presence during the times of necessity as an industry mentor to the pilot implementation project team.